\documentclass[11pt]{amsart}

\usepackage{xspace}
\usepackage{multirow}
\usepackage{graphicx}
\usepackage{hyperref}

\newcommand{\gene}{gene\xspace}
\newcommand{\transcript}{transcript\xspace}

\newcommand{\celltype}{cell type\xspace}
\newcommand{\subregion}{spatiotemporal subregion\xspace}

\title[Reconstructing Spatiotemporal Gene Expression Data]{Reconstructing Spatiotemporal Gene Expression Data from Partial Observations}
\author[DA Cartwright {\itshape et al.}]{Dustin A. Cartwright}
\address[Dustin A. Cartwright]{Department of Mathematics \\ University of California \\ Berkeley, CA}
\email{dustin@math.berkeley.edu}
\author[]{Siobhan M. Brady}
\address[Siobhan M. Brady]{Department of Plant Biology and Genome Center \\ University of California \\ Davis, CA}
\email{sbrady@ucdavis.edu}
\author[]{David A. Orlando}
\address[David A. Orlando]{Department of Biology and Program in Computational Biology and Bioinformatics \\ Duke University \\ Durham, NC}
\email{david.orlando@duke.edu}
\author[]{Bernd Sturmfels}
\address[Bernd Sturmfels]{Department of Mathematics \\ University of California \\ Berkeley, CA}
\email{bernd@math.berkeley.edu}
\author[]{Philip N. Benfey}
\address[Philip N. Benfey]{Department of Biology and Duke Center for Systems Biology \\ Duke University \\ Durham, NC}
\email{benfeyp@duke.edu}
\thanks{The second and third authors contributed equally to this work.}

\begin{document}

\begin{abstract}
Developmental transcriptional networks in plants and animals operate in both space and time. To understand these
transcriptional networks it is essential to obtain whole-genome expression data
at high spatiotemporal resolution.
Substantial amounts of spatial and temporal
microarray expression data previously have been obtained for the \textit{Arabidopsis}
root; however, these two dimensions of data have not been integrated
thoroughly. Complicating this integration is the fact that these
data are heterogeneous and incomplete, with observed expression levels representing
complex spatial or temporal mixtures. Given these partial observations, we
present a novel method for reconstructing integrated high resolution
spatiotemporal data. Our method is based on
a new iterative algorithm for finding
approximate roots to systems of bilinear equations.  
\end{abstract}

\maketitle

\section{Introduction}

Transcriptional regulation plays an important role in orchestrating a host of
biological processes, particularly during
development~(reviewed in~\cite{iyer-pascuzzi-benfey,levine2005grn}).
Advances in microarray and sequencing technologies have
allowed biologists to capture genome-wide \gene expression data; the output of
this transcriptional regulation. This expression data can then be used to identify \gene{}s
whose expression is correlated with a particular biological process,
and to identify transcriptional regulators that
coordinate the expression of groups of \gene{}s that are important for
the same biological process.

\begin{figure}[ht]
\includegraphics[scale=0.75]{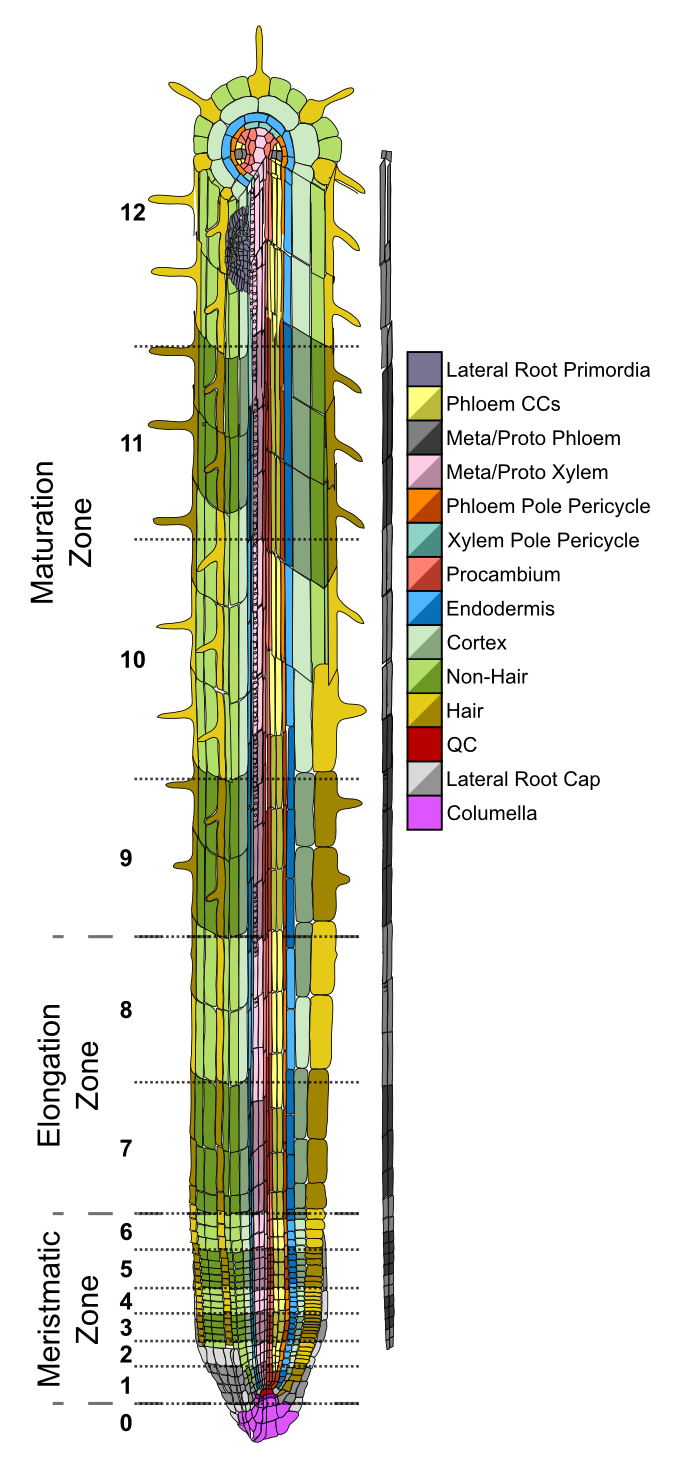}
\caption{\textit{Arabidopsis} root template used for expression pattern overlays.} \label{fig:template}
\end{figure}

The identification of such \gene{}s and transcriptional regulators is complicated
by the complex heterogeneous mixture of \celltype{}s and developmental stages
that comprise each organ of an organism. Expression patterns that are found only
in a subset of cell types within an organ will be
diluted and may not be detectable in the
collection of expression patterns obtained from RNA isolated from samples of an
entire organ.  Therefore techniques have been developed to enrich samples for
specific \celltype{}s or developmental stages,
especially for studies in
plants~\cite{busch2007}. In the model plant, {\itshape Arabidopsis thaliana}, several
features of the root organ reduce its developmental complexity and facilitate
analysis. Specifically, most root \celltype{}s are found within concentric cylinders moving from
the outside of the root to the inside of the root (Figure~\ref{fig:template}).  These cell type layers display rotational
symmetry thus simplifying the spatial features of development. This feature has
been exploited in the development of a \celltype enrichment method.  This
enrichment method uses
green fluorescent protein (GFP)-marked
transgenic lines and fluorescently-activated cell sorting (FACS) to collect
\celltype enriched samples and has allowed for the identification of
\celltype-specific expression patterns~\cite{birnbaum2005,birnbaum2003}. Using
this technique, high resolution expression data have been obtained
for nearly all \celltype{}s in
the \textit{Arabidopsis} root (herein called the \textit{marker-line
dataset})~\cite{brady,jiao}.

Another feature that makes the \textit{Arabidopsis} root a tractable
developmental model is that \celltype{}s are
constrained in files along the root's longitudinal axis and most of these cells are produced from a stem cell
population found at the apex of the root.  This feature allows a
cell's developmental timeline to be represented by its position along the length
of the root.  To obtain a developmental
time-series expression dataset individual
\textit{Arabidopsis} roots were sectioned into thirteen pieces, each piece
representing a developmental time point (herein called the
\textit{longitudinal
dataset})~\cite{brady}.  Each of these sections, however, contains a mixture of
\celltype{}s, and the microarray expression values obtained are therefore the average of
the expression levels over multiple \celltype{}s present at these
specific developmental time~points.

While the 19 fluorescently marked lines
in Brady {\itshape et al.\/}~\cite{brady} cover expression in nearly all \celltype{}s, they
do not comprehensively mark all developmental stages of these
\celltype{}s.  Also, the procambium \celltype was not measured, as a fluorescent
marker-line that marks that \celltype did not exist at the time.  However, expression from
the longitudinal dataset, does contain averaged expression of all
\celltype{}s, and may be used to infer the missing cell type data. 

Previous studies have looked at separating expression data from the
heterogeneous cell populations that make up tumors into the contributions of
their constituent cell types~\cite{ghosh,venet}. However, in that context, the
difficulty comes from the fact that the mixture of \celltype{}s in each sample
is unknown, whereas within our experimental context, the \celltype mixture of
each sample is known.
Two computational methods have been developed to combine the \textit{Arabidopsis} longitudinal and marker-line
datasets as experimentally resolving this expression with marker lines is nearly impossible~\cite{brady,chaudhuri2008}. 
However, neither method takes all
data into account when reconstructing expression. In~\cite{brady}, only high
relative gene expression is considered, and in~\cite{chaudhuri2008}, no attempt
is made to infer expression for cells not covered by any marker-line.

In this work we formulate a model for expression levels in \textit{Arabidopsis} roots in
which \celltype and developmental stage are independent sources of variation. 
The microarray data specifying overall expression levels for certain mixtures of
cells lead to an overconstrained system of bilinear equations.  Moreover, due
to the nature of the problem, we are exclusively interested in positive real
solutions. We present a new method for finding non-negative real approximate
solutions to bilinear equations, based on the techniques of expectation
maximization (EM)~\cite[Sec.\ 1.3]{ascb} and iterative proportional fitting
(IPF)~\cite{darroch-ratcliff} from likelihood
maximization in statistics. Earlier
work has used expectation maximization to find non-negative matrix
factorizations~\cite{lee-seung-alg}, and our method is a generalization of that
work.

We applied our method to estimate \subregion expression patterns for 20,872
\textit{Arabidopsis} transcripts. These patterns have identified gene expression in cell types
and developmental stages which were previously unknown.  A searchable database of verlays of these patterns on a schematic 
\textit{Arabidopsis} root 
is under development and will be made publicly available at
\url{http://www.arexdb.org}.

\section{Methods}

\subsection{Expression data}\label{subsec:expData}

Our method uses the normalized expression data collected in~\cite{brady}.
Expression levels were measured across 13 longitudinal sections in a single root (\textit{longitudinal
dataset}) and across 19 different markers (\textit{marker-line
dataset}). For simplicity, the J2501 line was removed from further
analysis as it is redundant with the WOODEN-LEG marker-line. The
APL marker-line was also removed, as it contains domains of expression marked by
both the S32 and the SUC2 marker-lines and adds no extra information. The
remaining 17 markers covering 14 cell types are listed in the second column
of Table~\ref{tbl:parameters}.

Due to computational constraints, the original normalization of
this data was performed for the longitudinal and the
marker-line datasets independently~\cite{brady}. In order to
account for differences caused by these separate normalization
procedures, we adjusted the marker-line data by a
global factor of $0.92$.
This factor was calculated by comparing the expression
values of ubiquitous, evenly expressed probe sets between the two
datasets. We assume that by comparing
these probe sets, any true expression
differences due to \celltype and longitudinal section specificity should be
minimal and thus any differences in expression level
is a byproduct of the separate
normalization procedures. A set of 43 probesets were identified which were
expressed ubiquitously (above a normalized value of 1.0 in all samples) and
whose expression did not vary significantly among samples within a dataset
(ratio of min/max expression within a dataset is at most~$0.5$). The scaling factor
necessary to make the mean expression within the
marker-line dataset equal to the mean expression
within the longitudinal dataset was calculated for each probe set in this set. The
median value of these 43 scaling factors was $0.92$, which was
used as the global adjustment factor (Table~\ref{tab:scalingTable}).

\subsection{Model} \label{subsec:model}

To model the \transcript
expression level of an individual cell we assume
that the effects of its \celltype and its section on its expression level are
independent of each other. More precisely, we assume that the transcript expression level
of a cell of type~$j$ in section~$i$ is equal to the product $x_i
\cdot y_j$, where $x_i$ depends only on the section and $y_j$
depends only on the \celltype{}. In other words, for each
\transcript, there is an idealized profile of expression
over different \celltype{}s, and an idealized profile of expression over
different sections.  Within a given section, our assumption is that
the \transcript expression level varies proportionally to
its \celltype profile, and within a given
\celltype{}, proportionally to its longitudinal
profile.

\begin{table}
\begin{tabular}{|l|l|l|}
\hline Cell type & Marker-lines \\
\hline
Quiescent center & AGL42, RM1000, SCR5 \\
Columella & PET111 \\
Lateral root cap & LRC \\
Hair cell & COBL9 (8-13) \\
Non-hair cell & GL2  \\
Cortex & J0571, CORTEX (7-13) \\
Endodermis & J0571, SCR5 \\
Xylem pole pericycle & WOL (2-9), JO121 (9-13), J2661 (13) \\
Phloem pole pericycle & WOL (2-9), S17 (8-13), J2661 (13) \\
Phloem & S32, WOL (2-9) \\
Phloem companion cells & SUC2 (10-19), WOL (2-9) \\
Xylem & S4 (2-7), S18 (8-13), WOL (2-9) \\
Lateral root primordia & RM1000 \\
Procambium & WOL (2-9) \\
\hline
\end{tabular}
\caption{The 14~\celltype{}s in the \textit{Arabidopsis} root and the
17~marker-lines which mark them~\cite{brady}. For markers that
only mark the cell type in some of the sections, these sections are indicated by
the range in parenthesis.} \label{tbl:parameters}
\end{table}

\begin{table}
\begin{equation*}
\begin{pmatrix}
0 & 24 & 51 & 0 & 0 & 0 & 0 & 0 & 0 & 0 & 0 & 0 & 0 & 0 \\
4 & 12 & 152 & 24 & 48 & 12 & 12 & 12 & 22 & 0 & 0 & 12 & 0 & 28 \\
0 & 0 & 280 & 40 & 80 & 40 & 40 & 20 & 45 & 20 & 20 & 25 & 0 & 80 \\
0 & 0 & 210 & 40 & 80 & 40 & 40 & 20 & 45 & 20 & 20 & 25 & 0 & 80 \\
0 & 0 & 210 & 40 & 80 & 40 & 40 & 20 & 45 & 20 & 20 & 25 & 0 & 80 \\
0 & 0 & 210 & 40 & 80 & 40 & 40 & 20 & 45 & 20 & 20 & 25 & 0 & 80 \\
0 & 0 & 0 & 40 & 80 & 40 & 40 & 20 & 45 & 20 & 20 & 25 & 0 & 80 \\
0 & 0 & 0 & 40 & 80 & 40 & 40 & 20 & 45 & 20 & 20 & 25 & 0 & 80 \\
0 & 0 & 0 & 40 & 80 & 40 & 40 & 20 & 45 & 20 & 20 & 25 & 0 & 80 \\
0 & 0 & 0 & 40 & 80 & 40 & 40 & 20 & 45 & 20 & 20 & 25 & 0 & 80 \\
0 & 0 & 0 & 40 & 80 & 40 & 40 & 20 & 45 & 20 & 20 & 25 & 0 & 80 \\
4 & 0 & 0 & 40 & 80 & 40 & 40 & 20 & 45 & 20 & 20 & 25 & 130 & 80 \\
0 & 0 & 0 & 40 & 80 & 40 & 40 & 20 & 45 & 20 & 20 & 25 & 0 & 80
\end{pmatrix}
\end{equation*}
\caption{The cell count matrix gives the number of cells in each \subregion. The
13 rows correspond to longitudinal
sections 1 through 13. From left to right, the 14 columns correspond to the
following \subregion{}s: quiescent center, columella, lateral root cap, hair cell, non-hair cell, cortex,
endodermis, xylem pole pericycle, phloem pole pericycle, phloem, phloem companion cells,
xylem, lateral root primordia, and procambium.} \label{tbl:cell-count}
\end{table}

Each microarray sample in the two datasets (described in
Section~\ref{subsec:expData}), is composed of a distinct mixture of
\celltype{}s and sections. Within each sample, the measured
\transcript expression level is a convex linear combination of the expression
levels of its constituent cells. Under the above assumptions, these
measurements constitute a system of bilinear equations,
\begin{equation} \label{eqn:model}
\sum_{i=1}^{13} \sum_{j=1}^{14} a_{ijk} x_i y_j = b_k \quad \mbox{for } k = 1,
\ldots, 30
\end{equation}
where $x_i$ and $y_j$ are the model parameters for the 13 sections and 14
\celltype{}s respectively, and $b_k$ is the measured expression level as $k$
ranges over the 30 measured samples (13 longitudinal sections and 17 markers).

The coefficients $a_{ijk}$ are obtained by combining the cell type by marker line data
(Table~\ref{tbl:parameters}) with the the cell-count matrix
(Table~\ref{tbl:cell-count}).
For $k$ at most $13$, the measurement with index $k$ comes
from the $k$th longitudinal
section. We will use $a_{**k}$ to denote the corresponding matrix, with
$a_{ijk}$ in the $i$th row and $j$th column.
We set this matrix to be zero everywhere except the $k$th
row, where it is proportional to the $k$th row of the cell-count matrix, but
rescaled to sum to~$1$. For $k$ greater than $13$, the measurements come from
one of the $17$~marker-lines. The matrix $a_{**k}$ is likewise zero except for
those
\subregion{}s marked by that marker as indicated in Table~\ref{tbl:parameters}.
Note that the non-zero entries of $a_{**k}$ may span multiple columns for those
markers which are listed in multiple rows of Table~\ref{tbl:parameters}. The
non-zero
entries of $a_{**k}$ are proportional to the corresponding entries of the cell
matrix, but rescaled to sum to~$1$.

\subsection{Cell matrix}\label{subsec:cellMatrix}

As described in the previous section, the coefficients $a_{ijk}$ in our model
depend on the number of cells in each \subregion. These cell number estimates
were generated by visual inspection of
successive optical cross-sections of \textit{Arabidopsis} roots along the
longitudinal axis using confocal laser scanning microscopy.  For the xylem,
phloem and procambium \celltype{}s, cell counts were obtained
from earlier experiments~\cite{bonke2003arv,mahonen2000ntc}. What follows is a detailed
description of this visual and literature analysis.  These results are also summarized in Table~2.

Longitudinal section 1 encompasses two tiers of 12 columella cells, and three
tiers of lateral root cap cells (15, 18 and 18 moving up from the tip).

Longitudinal section 2 contains one tier of 12 columella cells and six tiers of
lateral root cap cells (20, 20, 28, 28, 28 and 28 moving up from the tip).  For
all other \celltype{}s in longitudinal section~2,
three tiers of cells are present.  Eight trichoblast
(hair cell precursor) cells and 16 atrichoblast (non-hair cell precursor) cells
are present circumferentially throughout the root, resulting in 24 and 48 cells
respectively in the hair cell and non-hair cell precursor files in longitudinal
section 2. Throughout the root, eight cortex and
eight endodermis cells are present circumferentially.
However in longitudinal section~2, the cortex/endodermis initial is undergoing
assymetric periclinal divisions to produce the cortex and endodermis
cell files, so we consider there to be approximately 0.5 cells of the
cortex and endodermis type, resulting in 12 cells of each type in longitudinal
section~2. When the \textit{Arabidopsis} root is seven days old, each longitudinal section
from 3--13 contains approximately five cells of each
type along the root's longitudinal~axis.

In longitudinal section 2, the tangential and periclinal divisions that give
rise to phloem cell files do not occur, but do
occur in longitudinal section~3~\cite{bonke2003arv}.  Three cells are present in
the main xylem axis in the first tier of cells, four
cells in the second tier, and five cells in the third
tier~\cite{mahonen2000ntc}.  Eight procambial cells are present in the first
cell tier, 12 procambial cells in the second tier, and 18 cells in the third
tier resulting in 28 procambial cells in longitudinal
section~2~\cite{mahonen2000ntc}. For all sections xylem pole pericycle cells are
the two
cells that flank the xylem axis on either end, and phloem pole pericycle cells
are considered the intervening cells.  Four pericycle cells can be identified as
flanking xylem cells in all three tiers of cells
present in longitudinal section~2~\cite{mahonen2000ntc}.  Seven intervening
phloem pole pericycle cells can be found in tier one,
and eight
intervening cells can be identified in the third tier~\cite{mahonen2000ntc},
resulting in 22 procambial cells in longitudinal section~2.

In a seven day old root, each of the longitudinal sections 3--13 contains
approximately five tiers of cells.
In longitudinal section 3, columella cells can no longer be
identified, and 10 tiers of lateral root cap cells exist containing 28 cells
each.  In sections 4--6, a lateral root cap cell is 
twice the length and half the width of an epidermal cell.
Eighty-four cells were identified in each tier, and two and a half tiers of
cells exist each for longitudinal sections 4--6 resulting in 210 cells for each
longitudinal section.  All other \celltype{}s have undergone the appropriate
tangential and periclinal divisions to establish their respective cell files by
longitudinal section 3.  Two protophloem cells, two metaphloem cells and four
accompanying companion cells are present in the phloem
tissue~\cite{bonke2003arv}.  With the combination of protophloem and metaphloem
cells, 20 phloem cells and 20 companion cells exist in each longitudinal
section.  Approximately 40 procambial cells exist in each longitudinal section. 
Secondary cell growth does not occur in the developmental stages sampled,
therefore, this number remains fixed throughout all developmental stages.  In
longitudinal section 12, a non-emerged lateral root is hypothesized to be
present based on microarray expression data~\cite{brady}.  This lateral root is
estimated to be approximately 130 cells, or one tier of cells in
longitudinal section~2.

In our modelling the distinct vasculature, protophloem and metaphloem
\celltype{}s were treated as a single \celltype, as no marker-line was specific enough to differentiate
clearly between these \celltype{}s.  Also, the
metaxylem and protoxylem were considered as a single \celltype by
the same rationale.

\subsection{Solving Bilinear Equations}

In this section we present our method for solving the system of bilinear
equations given by~(\ref{eqn:model}). More generally, we have a system
\begin{equation} \label{eqn:bilinear}
f(x,y) := \sum_{i = 1}^n \sum_{j=1}^m a_{ijk} x_i y_j = b_k
\quad\mbox{for }k = 1, \ldots,\ell.
\end{equation}
In our application, we have $n=13$, $m=14$, and $\ell=30$.
Unlike other numerical methods for solving systems of polynomial equations,
our algorithm has the advantage that it finds only non-negative, real solutions.
Moreover, even in systems where there are no exact solutions, as will generally
be the case in an overconstrained system of equations, our method will find
approximate solutions. A more detailed, technical mathematical study of the method will be
available in a forthcoming paper by the first author.

Our method is based on the Expectation-Maximization (EM)~\cite[Sec.\ 1.3]{ascb} and Iterative
Proportional Fitting (IPF)~\cite{darroch-ratcliff} algorithms used for maximum
likelihood estimation in
statistics.  These are iterative
algorithms which reduce the modified Kullback-Leibler divergence at each
step:
\begin{equation} \label{eqn:generalized-kl}
D(f(x,y) \Vert b) = \sum_{k=1}^\ell \left(b_k \log\left(\frac{b_k}{f_k(x,y)}\right) -
b_k + f_k(x,y) \right).
\end{equation}
The traditional Kullback-Leibler divergence consists only of the first
term in the summation. The other two terms are a natural
generalization, which is necessary only when the vectors
$f(x,y)$ and $b$ do not sum to one~\cite{lee-seung-alg}.

Our algorithm begins with an arbitrarily chosen starting point 
$(x^{(0)}, y^{(0)})$ in $\mathbb R_{> 0}^{m + n}$.
In each iteration $s$, the expectation step computes the quantities:
\begin{equation}\label{eqn:E-step}
w_{ijk}^{(s)} := b_k \frac{a_{ijk} x_i^{(s)} y_j^{(s)}}
{\sum_{i'=1}^n \sum_{j'=1}^m a_{i'j'k} x_{i'}^{(s)} y_{j'}^{(s)}} 
\end{equation}
for all $i$, $j$, and~$k$.
This quantity $w_{ijk}^{(s)}$ is an estimate of the contribution of the $(i,j)$
term in the $k$th equation in~(\ref{eqn:bilinear}).
The maximization step is an analogue of the IPF algorithm, and itself consists
of an iteration. We first compute the analogues of the sufficient statistics:
\begin{align*}
X^{(s)}_i &= \sum_{j = 1}^m \sum_{k=1}^\ell w_{ijk}^{(s)} \\
Y^{(s)}_j &= \sum_{i = 1}^n \sum_{k=1}^\ell w_{ijk}^{(s)}.
\end{align*}
Then we perform an iteration beginning with $x_i^{(s,0)} = x_i^{(s)}$ and
$y_j^{(s,0)} = y_j^{(s)}$ and
the update rules
\begin{align*}
x_i^{(s,t+1)} &:= x_i^{(s,t)} \frac{X_i^{(s)}}{\sum_{j = 1}^m \sum_{k=1}^\ell a_{ijk} x_i^{(s,t)} y_j^{(s,t)}} \\
y_j^{(s,t+1)} &:= y_j^{(s,t)} \frac{Y_j^{(s)}}{\sum_{i = 1}^n \sum_{k=1}^\ell
a_{ijk} x_i^{(s,t+1)} y_j^{(s,t)}}
\end{align*}
until the parameters converge. We then re-normalize and use the values from the
last index~$t$ for the next
step of the EM algorithm: 
\begin{align*}
x_i^{(s+1)} &:= \frac{x_i^{(s,t)}}{\sum_{i' = 1}^n x_{i'}^{(s,t)}} \\
y_j^{(s+1)} &:= y_j^{(s,t)}\sum_{i' = 1}^m x_{i'}^{(s,t)}.
\end{align*}

At each step of each of these algorithms,
the Kullback-Leibler divergence defined in~(\ref{eqn:generalized-kl}) decreases. In the statistics
literature, the convergence of the EM and IPF algorithms is known
under the additional assumptions that
\begin{equation*}
\sum_{i=1}^n x_i = \sum_{j=1}^m y_j = \sum_{k=1}^\ell b_k = 1
\end{equation*}
However, relaxing these conditions does not change the convergence proof.

We repeatedly ran our EM algorithm beginning with $20$ different randomly chosen
starting points. For each transcript in the data, all $20$ runs of the algorithm
converged to the same solution, strongly suggesting that we have found a
global minimum to the modified Kullback-Leibler divergence.

\subsection{Computational validation methodology}
In order to validate our method, we simulated expression profiles according to
various models and tested our method's ability to reconstruct
the underlying parameters. First, we simulated data according to
the same independence model defined in Section~\ref{subsec:model}. The
underlying
\subregion expression levels were sampled from a log-normal distribution with
standard deviation~$0.5$. The simulated measurements~$b_k$
were computed from these subregion levels according to our
model of the \textit{Arabidopsis} root in (\ref{eqn:model}). Finally, multiplicative error was added,
distributed according to a log-normal distribution with standard
deviation~$0.03$ to simulate measurement noise. This procedure created
expression data with varying but comparable expression levels, which we will
call the ``uniform'' dataset. However, since we are particularly interested in
genes for which the expression levels are not uniform, we
also produced simulations with the expression level for a given
section or \celltype raised by a factor of~$10$, which we will call the
``elevated'' dataset. In this dataset, we only measured the error for the same
section or \celltype which was elevated.
These simulations measure our ability to detect a dominant
expression pattern.

In addition, we designed simulations that test the robustness of the algorithm
to failures of the bilinear model for root expression levels. For each section
and \celltype{}, we simulated data in which the expression levels for cells in
that section or \celltype did not follow the bilinear model, and call
these the ``section'' and ``cell type'' datasets respectively. Instead, the
expression levels in the given section or \celltype were
chosen independently according to a log-normal distribution with standard
deviation~$0.5 \sqrt{2}$.
The factor of $\sqrt{2}$ was introduced because
the product of two log-normally distributed numbers with standard deviation
$0.5$ is distributed log-normally with standard deviation $0.5\sqrt{2}$.

The predictions were compared to the true expression levels across the
\subregion{}s within each section and each \celltype. For each section and each
\celltype, the expression levels in its \subregion{}s were averaged, ignoring
those combinations which are not physically present in the root, (i.e.\ those
whose entry in Table~\ref{tbl:cell-count} is $0$). The
difference between the predicted and true average expressions was computed as a
proportion of the true average expression. We then computed the root mean square
of the proportional error over $500$ simulations.

\subsection{Visualization of predicted expression patterns}

Predicted expression values were colored according to an
\textit{Arabidopsis} root template (Figure~\ref{fig:template}). The green channel of each
cell was set according to a linear mapping between the expression range shown in
the template $[1,10]$ or $[1,5]$ to the range $[0,255]$. Expression values above
or below that range are given values of $255$ or $0$ respectively. The mapping
is
also shown to the right of the false color image in the form of a
gradient key. Phloem cells by longitudinal section are visualized separately on
the right hand side of the root as they are physically occluded by other 
cells in the left hand side representation. The minimum and maximum range of expression value visualized can also be adjusted by the user.

\subsection{{\itshape In vivo\/} validation methodology}

To validate predicted expression values, we used
transgenic \textit{Arabidopsis thaliana} lines containing transcriptional       
GFP fusions in the Columbia ecotype~\cite{lee2006tap}.
For each gene being validated, six plants from at least two insertion lines
previously described as expressing GFP were characterized.  All plants were     
grown vertically on 1X Murashige and Skoog salt mixture, 1\% sucrose and 2.3     
mM 2-(\textit{N}-morpholino)ethanesulfonic acid (pH 5.7) in 1\% agar.             
Seedlings were prepared for microscopy at 5 days of age.  Confocal images       
were obtained using a 25x water-immersion lens on a Zeiss LSM-510 confocal      
laser-scanning microscope using the 488-nm laser for excitation.  Roots    
were stained with 10 $\mu$g/mL propidium iodide for 0.5 to 2 minutes and mounted    
in water.  GFP was rendered in green and propidium iodide in red.  Images       
were saved in TIFF format.  Images were manually stitched together in Adobe     
Photoshop CS2 using the Photomerge command.  The black background               
surrounding the root was modified to ensure uniformity across figures.  No      
other image enhancement was performed.                                          

\section{Results}

\begin{table}
\begin{tabular}{|l|r|r|r|r|}
\hline
& \multicolumn{4}{c|}{Error rate} \\
Variable & uniform & elevated & \celltype & section \\
\hline
Section 1 & 2.7 & 2.4 & 3.3 & 3.6 \\
Section 2 & 3.4 & 3.0 & 5.7 & 7.5 \\
Section 3 & 3.3 & 2.7 & 5.8 & 7.2 \\
Section 4 & 3.2 & 2.8 & 5.3 & 6.5 \\
Section 5 & 3.1 & 2.7 & 5.3 & 6.5 \\
Section 6 & 3.3 & 2.7 & 5.3 & 6.5 \\
Section 7 & 3.1 & 2.5 & 3.7 & 5.0 \\
Section 8 & 3.0 & 2.3 & 3.6 & 4.9 \\
Section 9 & 3.0 & 2.2 & 3.6 & 4.8 \\
Section 10 & 2.7 & 2.1 & 3.5 & 4.5 \\
Section 11 & 2.9 & 2.2 & 3.4 & 4.6 \\
Section 12 & 3.3 & 2.2 & 4.4 & 5.3 \\
Section 13 & 2.4 & 2.1 & 3.6 & 5.3 \\
Quiescent center & 3.0 & 3.1 & 3.0 & 3.1 \\
Columella & 3.1 & 3.8 & 4.9 & 4.1 \\
Lateral root cap & 2.6 & 1.6 & 3.6 & 3.1 \\
Hair cell & 3.4 & 2.8 & 9.1 & 4.3 \\
Non-hair cell & 3.0 & 2.1 & 3.1 & 3.0 \\
Cortex & 2.9 & 2.1 & 6.9 & 3.6 \\
Endodermis & 2.8 & 2.2 & 3.5 & 3.2 \\
Xylem pole pericycle & 3.3 & 3.1 & 10.8 & 4.9 \\
Phloem pole pericycle & 3.0 & 2.9 & 9.4 & 4.9 \\
Phloem & 3.0 & 2.9 & 3.0 & 3.0 \\
Phloem ccs & 3.3 & 3.4 & 11.7 & 4.9 \\
Xylem & 2.2 & 2.1 & 2.5 & 2.2 \\
Lateral root primordia & 3.5 & 3.0 & 3.4 & 3.3 \\
Procambium & 8.3 & 1.8 & 12.7 & 12.7 \\
\hline
\end{tabular} 
\caption{Root mean square percentage error rates in the reconstruction of
simulated data. The first column is under a model of comparable but varying
expression levels across all sections and \celltype{}s. The second
type is the error rate when that section or \celltype has its
expression level raised by a factor of~$10$. The third and fourth
columns show models in which the bilinear assumption is violated in
one of the sections or one of the \celltype{}s respectively. In all
cases, $3\%$ measurement error has been added to the expression
levels.}
\label{tbl:computational-val}
\end{table}

\subsection{Computation validation}

The root mean square percentage errors in the reconstruction of each parameter
are shown in Table~\ref{tbl:computational-val}. In the first two columns, where
the data were
generated according to the bilinear
model, the error rate is generally no
greater than the simulated measurement error. In most cases, elevated expression
led to a lower error rate. In particular, reconstruction of
expression in procambium was much more accurate in the elevated dataset.

The last two columns show that the algorithm is robust to violations
of the bilinear model. Also, the predicted
expression level in each \celltype is generally not greatly affected by the
failure of the model in other \celltype{}s, and similarly with
sections.

\subsection{{\itshape In vivo\/}\ validation}

To determine whether our algorithm is able to accurately resolve
\subregion-level transcript expression values, it would be ideal to compare the
predictions to measured microarray expression values of the same \subregion. However, due to technical constraints, it is not
possible to measure mRNA expression to such a degree of specificity and thus we cannot
validate the estimates directly. Instead, we validated the
method by visually comparing the predicted pattern of expression to patterns
obtained from transcriptional GFP fusions using laser scanning
confocal microscopy, as described in~\cite{lee2006tap}.  

For each
gene validated, a false-colored root image was generated by coloring each
\subregion of an annotated {\it Arabidopsis\/} root template
(Figure~\ref{fig:template})
according to the
expression level in that subregion as predicted by our method. This
false-colored image was then visually compared against the actual pattern of
fluorescence observed in plants expressing a
transcriptional GFP fusion specific for the promoter of that \gene. These
transcriptional GFP fusions contain up to 3 kb of
regulatory sequence upstream of the translational start site of the respective
\gene. In many cases, this sequence
is sufficient to recapitulate endogenous mRNA expression
patterns as defined by \celltype resolution microarray data~\cite{lee2006tap}.
This comparative method of validation allows us to assess the
accuracy of \subregion expression predictions in
an efficient and technically feasible way.

\begin{figure}[ht]
\includegraphics[width=0.9\textwidth]{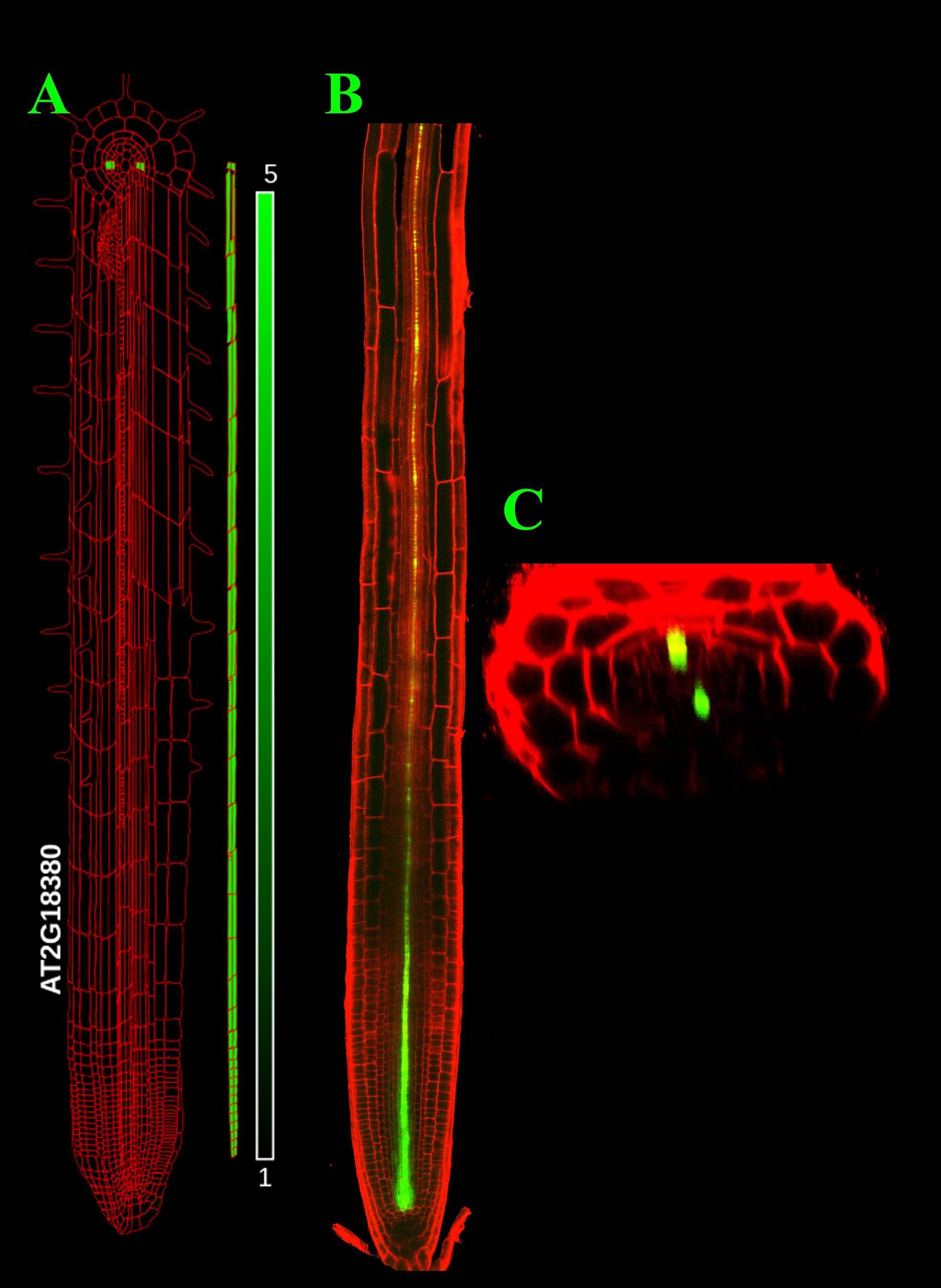}
\caption{(A) Expression of \textit{AT2G18380} in all developmental stages of
the phloem was predicted by our method and visualized in a
representation of the \textit{Arabidopsis} root.  Phloem cells are shown
external to the root.  (B) GFP expression in the longitudinal axis and
(C) expression in cross-section of expression driven by the \textit{AT2G18380}
promoter validate the prediction.} \label{fig:expr-phloem}
\end{figure}

As a benchmark validation test, a set of three transcriptional fusions which
were used to obtain some of the 
marker-line dataset were examined:
S18(\textit{AT5G12870}), S4(\textit{AT3G25710}), and S32(\textit{AT2G18380}). 
These fusions were originally selected for use in profiling
because they exhibited enriched \celltype expression as observed by laser
scanning confocal
microscopy and subsequently confirmed in the microarray expression data.
The expression predictions from our
method accurately recapitulated the observed pattern of all three
benchmark \gene{}s (Figure~\ref{fig:expr-phloem} and data not shown).

\begin{figure}[ht]
%Figure x.  
\includegraphics[width=0.9\textwidth]{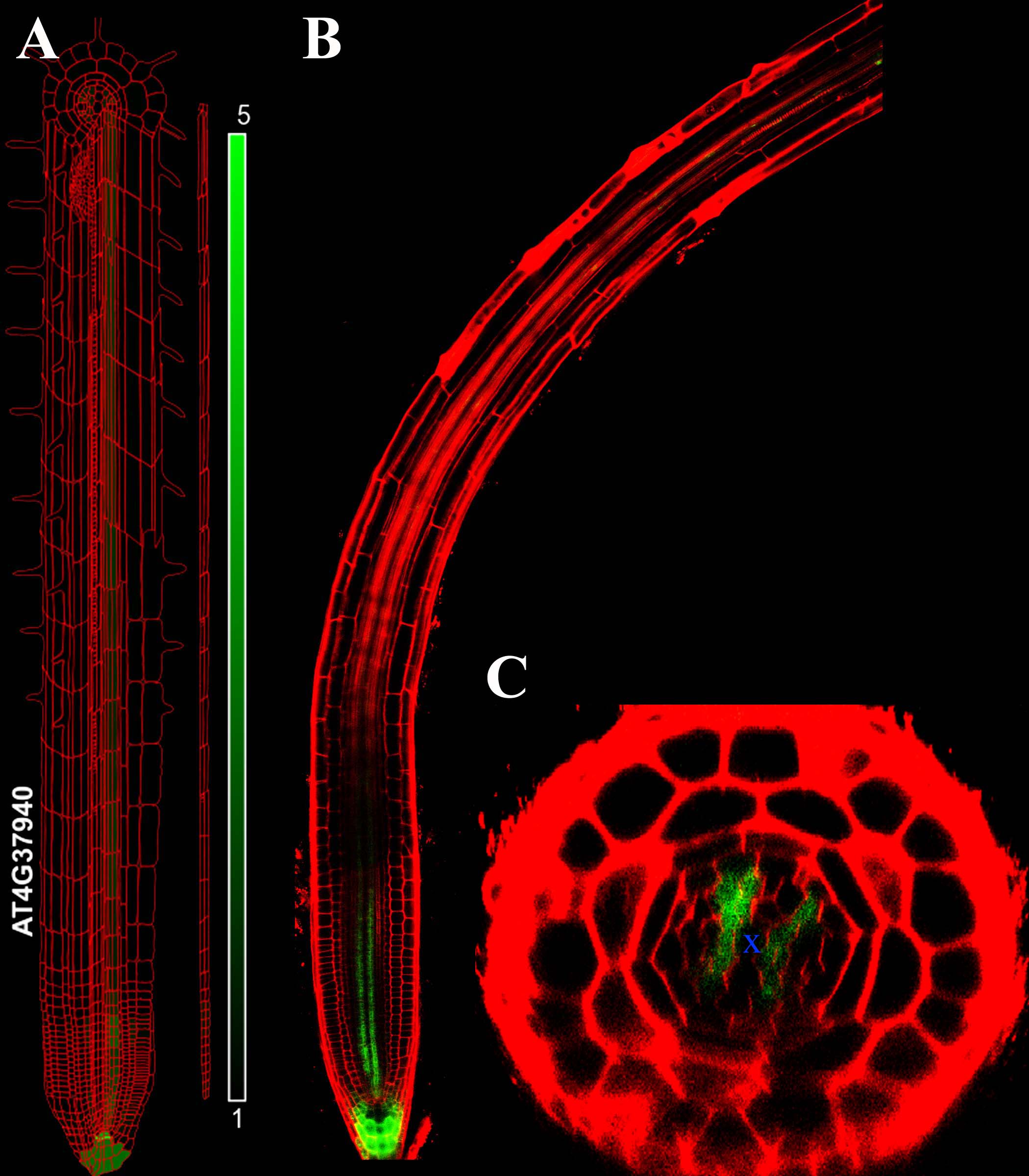}
\caption{(A) Our method correctly predicts specific expression of
a candidate gene in a \celltype{}, procambium, that is only covered by a
general tissue marker, WOL.  Expression conferred by the candidate gene
promoter fused to GFP as a reporter was visualized in the columella
(B) and in the procambium by a longitudinal section (B)
and a cross section~(C).  The label~X indicates the xylem axis.  The expression
also validates a maximal peak in the meristematic zone.}
\label{fig:expr-procambium}
\end{figure}

To assess the novel predictive ability of our 
method to reconstruct {\itshape in vivo\/} expression
patterns given missing data, we selected transcriptional fusions for \gene{}s
for which our method predicts expression
in \celltype{}s or in \subregion{}s that were not marked by fluorescent
marker-lines in the original dataset.  At least two lines per transcriptional fusion were monitored.  With
respect to an unmarked \celltype{}, we selected a candidate gene predicted be our model to be highly expressed in the columella
and developing procambium.  Imaging
of a transcriptional fusion of this \gene confirmed this
expression
(Figure~\ref{fig:expr-procambium}).

\begin{figure}[ht]
%Suppl. Figure x. 
\includegraphics[width=0.8\textwidth]{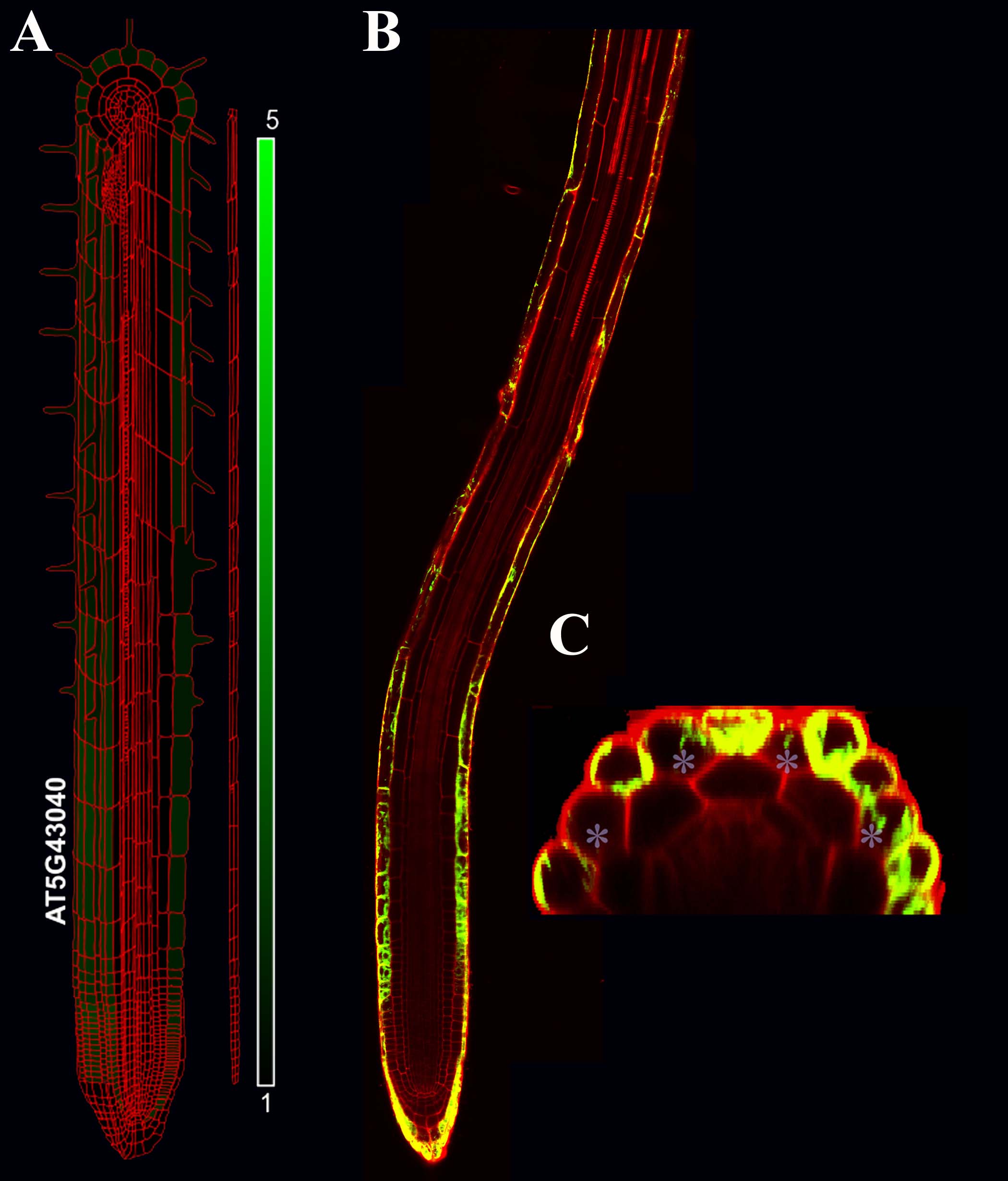}
\caption{(A) Our method correctly predicts candidate gene expression in trichoblast cells in the meristematic zone, which are not
currently covered by any marker-lines.  Furthermore, the algorithm was
able to predict differential expression within epidermal tissue with
high expression in non-hair cells and atrichoblasts (immature non-hair
cells in the meristematic and elongation zone) and decreased
expression in hair cells or trichoblasts (immature hair cells in the
meristematic and elongation zone).  Expression conferred by the
candidate gene promoter fused GFP as a reporter was visualized in the
epidermis in a longitudinal section (B) and was specifically
identified as high in atrichoblasts, and lower in trichoblasts (marked
with an asterisk) in cross-section (C).  Trichoblasts or hair cells
differentiate at the junction between two underlying cortical cells.}
\label{fig:expr-trich}
\end{figure}

We next determined if our method could
correctly differentiate expression in a specific developmental stage of a
\celltype{}.  The collection of marker-lines
used to generate the
original dataset included a
marker for all developmental stages of non-hair cells,
composed of their precursors (atrichoblasts)
and fully developed non-hair cells.
However, the marker-line used for hair cells only marks mature hair cells, and not their precursors (trichoblasts).
Using predictions from our method we tested a candidate gene with predicted
expression throughout the epidermis---in mature hair cell, 
trichoblast, mature non-hair cell and atrichoblast cell files---with higher expression predicted in
non-hair cells than in hair cells.  This differential expression was validated
using a transcriptional fusion (Figure~\ref{fig:expr-trich}) 
demonstrating that our method is not only able
to identify expression in a developmental stage of a \celltype not marked by the
marker-line data, but also to accurately differentiate relative
levels of a \transcript. However, it should be noted that expression in the
transcriptional fusion did not fully corroborate the expression predicted by
our algorithm---specifically, expression was found in the lateral root cap
which was not predicted by our algorithm. 

Examination of the raw microarray
expression data revealed that expression was not elevated in the lateral root
cap in the input microarray data. 
Most likely, the presence of GFP is not indicative of erroneous reconstruction of gene
expression in this case.  Instead, the transcriptional fusion does not contain
sufficient regulatory elements to direct the appropriate expression as described in~\cite{lee2006tap}, perhaps 
within downstream sequences.  For this reason, a comparison of the ratio between raw marker line and 
section expression data can be obtained as a link for each gene so that the user can simultaneously assess
raw expression data with the reconstructed expression patterns.

\section{Discussion}

We have shown that spatiotemporal patterns of gene expression in the
\textit{Arabidopsis} root can be reconstructed using information from the marker-line and
longitudinal datasets. Current experimental techniques are limited in their
ability to rapidly and accurately microdissect organs into all component cell types
at all developmental stages. Our computational technique helps to
overcome these limitations. We fully integrate the marker-line and longitudinal data
sets into a comprehensive expression pattern, across both space and time.  In particular, 
this method has enabled the identification of \textit{Arabidopsis} root procambium and trichoblast-specific genes, which have 
been previously experimentally intractable cell types.

Our high-resolution expression patterns will allow us to better understand the
regulatory logic that controls developmental processes of the \textit{Arabidopsis} root.
These transcriptional regulatory networks are key to understanding developmental
processes and environmental responses.
With only a portion of these genes and fewer \celltype{}s, high-resolution
spatiotemporal data
has been used to identify
transcriptional regulatory modules~\cite{brady}. Our more accurate and complete
dataset will allow a more comprehensive discovery of regulatory networks across additional cell
types.

Moreover, we expect that our algorithm and the model which underlies it are applicable to time course
experiments on other heterogeneous cell mixtures.
Measurements in multicellular organisms are taken from
complex cell mixtures of organs, tissues, heterogeneous cell lines, or cancerous samples.
When precise histological characterization of these samples can estimate
underlying cell type composition,
our method can be used to reconstruct the underlying \celltype{}-specific gene expression patterns or any other 
type of quantitative data, such as high-throughput protein abundance measurements. Theoretically, this algorithm
can be applied to identify missing data in any experimental system that captures
data in two or more dimensions which
are assumed to be independent of one another.

\section{Acknowledgments}

The authors acknowledge the support of the DARPA project ``Microstates to Macrodynamics: A New Mathematics of Biology.'' 
DAC and BS were supported by the U.S.~National Science Foundation 
(DMS-0354321, DMS-0456960, and DMS-0757236).  SMB was an NSERC postdoctoral
scholar.  SMB, DAO and PNB acknowledge
support through the NSF AT2010 program.  The authors would like to thank
Wolfgang Busch, Anjali Iyer-Pascuzzi, 
Terri Long, Anne Shiu, and Rossangela Sozzani for critical review of the manuscript.

\bibliographystyle{plain}
\bibliography{bilinear}

\begin{table}
\label{tab:scalingTable}
\begin{tabular}{|l|l|l|l|l|}
\hline
Probe & Gene(s) & Longitudinal & Marker-line  & Ratio \\
\hline
246630\_at & AT1G50730 & 1.073 & 1.239 & 0.866 \\
246980\_at & AT5G67530 & 1.676 & 1.605 & 1.045 \\
247163\_at & AT5G65685 & 1.423 & 1.508 & 0.944 \\
247286\_at & AT5G64280 & 1.146 & 1.292 & 0.887 \\
247334\_at & AT5G63610 & 1.75 & 1.497 & 1.169 \\
247363\_at & AT5G63200 & 1.225 & 1.305 & 0.938 \\
248266\_at & AT5G53440 & 2.514 & 1.867 & 1.347 \\
250524\_at & AT5G08520 & 1.691 & 1.54 & 1.099 \\
250791\_at & AT5G05610 & 1.151 & 1.334 & 0.863 \\
251104\_at & AT5G01720;AT5G01715 & 1.214 & 1.257 & 0.966 \\
251233\_at & AT3G62800 & 1.171 & 1.204 & 0.973 \\
252157\_at & AT3G50430 & 1.398 & 1.21 & 1.155 \\
253409\_at & AT4G32960 & 1.406 & 1.572 & 0.894 \\
253565\_at & AT4G31200 & 1.461 & 1.326 & 1.101 \\
253826\_s\_at & AT4G27960;AT5G53300 & 26.884 & 26.78 & 1.004 \\
255253\_at & AT4G05000 & 1.167 & 1.696 & 0.688 \\
255704\_at & AT4G00170 & 1.35 & 1.853 & 0.729 \\
255725\_at & AT1G25540 & 1.507 & 1.636 & 0.921 \\
255838\_at & AT2G33490 & 1.691 & 1.473 & 1.148 \\
255946\_at & AT1G22020 & 1.335 & 1.481 & 0.901 \\
256236\_at & AT3G12350 & 1.293 & 1.978 & 0.654 \\
256907\_at & AT3G24030 & 1.601 & 1.751 & 0.915 \\
256961\_at & AT3G13445 & 1.383 & 1.441 & 0.96 \\
258269\_at & AT3G15690 & 1.794 & 1.489 & 1.205 \\
259243\_at & AT3G07565 & 1.678 & 1.768 & 0.949 \\
259280\_at & AT3G01150 & 1.302 & 1.941 & 0.671 \\
259313\_at & AT3G05090 & 1.275 & 1.642 & 0.776 \\
259341\_at & AT3G03740 & 1.342 & 1.606 & 0.835 \\
259800\_at & AT1G72175 & 1.159 & 1.4 & 0.828 \\
260133\_at & AT1G66340 & 1.278 & 1.636 & 0.781 \\
261348\_at & AT1G79810 & 1.16 & 1.464 & 0.792 \\
261515\_at & AT1G71800 & 1.236 & 1.366 & 0.905 \\
261634\_at & AT1G49970 & 1.468 & 1.732 & 0.847 \\
261666\_at & AT1G18440 & 1.258 & 1.297 & 0.97 \\
261744\_at & AT1G08490 & 1.265 & 1.384 & 0.914 \\
262089\_s\_at & AT1G56000;AT1G55980 & 1.554 & 1.194 & 1.301 \\
262379\_at & AT1G73020 & 1.15 & 1.398 & 0.823 \\
262672\_at & AT1G76050 & 1.306 & 1.356 & 0.963 \\
262860\_at & AT1G64810 & 1.412 & 1.39 & 1.016 \\
263984\_at & AT2G42670 & 1.327 & 1.314 & 1.01 \\
264307\_at & AT1G61900 & 1.273 & 1.718 & 0.741 \\
265129\_at & AT1G30970 & 1.333 & 1.411 & 0.945 \\
267401\_at & AT2G26210 & 1.198 & 1.476 & 0.812 \\
\hline
\end{tabular}
\caption{Mean expression values and scaling factors of ubiquitously, evenly
expressed probesets across longitudinal and marker-lines}
\end{table}

\end{document}